\documentclass[sigconf]{acmart}

\AtBeginDocument{%
  \providecommand\BibTeX{{%
    \normalfont B\kern-0.5em{\scshape i\kern-0.25em b}\kern-0.8em\TeX}}}

\copyrightyear{2023}
\acmYear{2023}
\setcopyright{acmlicensed}
\acmConference[WWW '23]{Proceedings of the ACM Web Conference 2023}{May 1--5, 2023}{Austin, TX, USA}
\acmBooktitle{Proceedings of the ACM Web Conference 2023 (WWW '23), May 1--5, 2023, Austin, TX, USA}
\acmPrice{15.00}
\acmDOI{10.1145/3543507.3583360}
\acmISBN{978-1-4503-9416-1/23/04}

\usepackage{graphicx}
\usepackage{amsfonts}
\usepackage{mathtools}
\usepackage{subcaption}
\usepackage{wrapfig}

\usepackage{makecell}
\usepackage{multirow}

\usepackage{enumitem}

\usepackage{xcolor} 
\usepackage{soul}

\begin{document}

\title{Learning To Rank Resources with GNN}


\author{Ulugbek Ergashev}
\affiliation{%
  \institution{Binghamton University}
  \city{Binghamton, NY}
  \country{USA}
  }
\email{uergash1@binghamton.edu}

\author{Eduard C. Dragut}
\affiliation{%
  \institution{Temple University}
  \city{Philadelphia, PA}
  \country{USA}
  }
\email{edragut@temple.edu}

\author{Weiyi Meng}
\affiliation{
  \institution{Binghamton University}
  \city{Binghamton, NY}
  \country{USA}
  }
\email{meng@binghamton.edu}


\begin{abstract}

As the content on the Internet continues to grow, many new dynamically changing and heterogeneous sources of data constantly emerge. A conventional search engine cannot crawl and index at the same pace as the expansion of the Internet. Moreover, a large portion of the data on the Internet is not accessible to traditional search engines. Distributed Information Retrieval (DIR) is a viable solution to this as it integrates multiple shards (resources) and provides a unified access to them. Resource selection is a key component of DIR systems. There is a rich body of literature on resource selection approaches for DIR. A key limitation of the existing approaches is that they primarily use term-based statistical features and do not generally model resource-query and resource-resource relationships. In this paper, we propose a graph neural network (GNN) based approach to learning-to-rank that is capable of modeling resource-query and resource-resource relationships. Specifically, we utilize a pre-trained language model (PTLM) to obtain semantic information from queries and resources. Then, we explicitly build a heterogeneous graph to preserve structural information of query-resource relationships and employ GNN to extract structural information. In addition, the heterogeneous graph is enriched with resource-resource type of edges to further enhance the ranking accuracy. Extensive experiments on benchmark datasets show that our proposed approach is highly effective in resource selection. Our method outperforms the state-of-the-art by 6.4\% to 42\% on various performance metrics.

\end{abstract}

\ccsdesc[500]{Information systems~Learning to rank}
\ccsdesc[500]{Information systems~Combination, fusion and federated search}

\keywords{selective search, resource selection, federated search}

\maketitle

\section{Introduction}
With the World Wide Web playing an increasingly central role in our daily lives, it is constantly growing with an overwhelming amount of heterogeneous and distributed information sources. A conventional web search engine or a centralized information retrieval system cannot explore and index web documents at the same rate that they change \cite{r:44,r:45}. Distributed Information Retrieval (DIR) \cite{r:7}, also known as Federated Search (FS) \cite{r:4, r:63, r:64, r:65}, has received much attention over the years as an effective solution to this problem.
DIR aggregates distributed information sources (resources) under a single query interface. A basic DIR system consists of multiple resources and a broker. When a broker receives a query from a user, it passes the query to a subset of carefully selected resources to satisfy the user's information need \cite{r:62, r:66}. After a query is processed by the selected resources, a ranked list of documents from each resource are retrieved to the broker \cite{r:67}. Finally, the broker merges the obtained results into a unified ranked list for presentation to the user \cite{r:58, r:59, r:60}. This paper focuses on resource representation and selection in DIR.

A rich line of methods exists for the resource representation and selection problems. Recent years have seen a great interest of application of learning to rank algorithms to the resource selection problem \cite{r:20,r:31,r:19}. However, these approaches do not consider fine-grained semantic information from both queries and resources, and are mainly limited to the statistical information like term and document frequencies of the sampled documents available in a Centralized Sample Index (CSI) \cite{r:1}. Moreover, they do not consider two pieces of important information: query-resource and resource-resource relationships. Some resources may have similar or even the same documents. Those resources should have a better chance of being relevant for the same query. One such approach uses a joint probabilistic classification model for resource selection \cite{r:28}. The model utilizes logistic regression to estimate the probability of relevance of information sources. While this approach is still one of the flagship solutions to the problem, it predates deep learning era and has a number of limitations, including that logistic regression requires no multicollinearity between independent variables, and its sensitivity to outliers \cite{r:33}.

In this paper, we propose a learning to rank based resource selection algorithm, named FedGNN. Learning to rank has attracted lots of research interest for ranking documents \cite{r:48,r:49,r:50,r:51}, however, its application to ranking resources has not been studied in depth \cite{r:19, r:28, r:20}. Resource selection is tightly coupled with the resource representation problem \cite{r:4,r:20}.
Hence, in this paper, we propose a solution that addresses both of these problems. Specifically, we utilize a pre-trained language model (PTLM) \cite{r:36} to extract semantic representations (embeddings) from query and resource data. Then, we leverage GNN to capture relationships between queries and resources. 
A GNN-based approach facilitates generalization across graph which allows the model to provide an appropriate response to a new query. This inductive capability of GNN is essential for high-throughput DIR systems, which operate on evolving graphs and constantly encounter unseen queries \cite{r:35}. 

This paper has the following contributions:
\begin{itemize}[leftmargin=*]
    \item To our knowledge, this is the first study to model a whole corpus as a heterogeneous graph for the resource selection problem. We leverage GNN that enriches embeddings with structural information from a heterogeneous graph.
    \item Our model extracts both (hidden) textual and structural features. 
    Different from other studies, we represent queries and resources as embeddings from a pre-trained language model, which help captures richer semantic information.
    \item The empirical results and analysis demonstrate that our methods outperform state-of-the-art resource selection methods by a significant margin. Our best method outperforms the best performing baselines by 6.4\% to 42\% on various performance metrics. 
\end{itemize}

The rest of this paper is organized as follows. Section \ref{sec:related-work} discusses related work. Section \ref{sec:method} presents the details of the proposed method. Section \ref{sec:experiments} describes and discusses the experimental results. Conclusion to this study and future work are given in Section \ref{sec:conclusions}.

\section{Related Work}
\label{sec:related-work}
We give an overview of the related literature in this section. We group existing resource selection methods broadly into four groups: lexicon-based, sample-based, supervised, and combination-based methods \cite{r:23,r:4,r:46}.

Lexicon-based methods
treat each resource as a big bag of words, and they collect word statistics from all documents to describe a resource \cite{r:7, r:8}. These approaches calculate the similarity between user query and each resource based on matching term frequencies, adapting traditional document ranking methods \cite{r:9}. 
This group of approaches has the limitation of not being able to encode semantics both at word and sentence granularity, which reduces the resource representation quality \cite{r:10}. 

Sample-based algorithms 
sample documents from each resource and build a Centralized Sample Index (CSI). Then, a user query is issued to the CSI to retrieve relevant documents, and each relevant document sampled from a resource is considered as a vote for the resource. Many methods of in this group, such as ReDDE \cite{r:11}, CRCS \cite{r:15} and SUSHI \cite{r:16}, utilize various voting algorithms over documents to infer a ranking of the resources. 
KBCS \cite{r:10}, the most recent method in this group, represents resources as a weighted entity set based on context- and structure-based measures, and ranks resources according to a query-resource similarity metric. Both lexicon-based and sample-based methods have been proved effective when the resources are mostly homogeneous \cite{r:23}, but they cannot fully utilize structural features of text inputs. 

Supervised methods employ machine learning approaches to build a model to tackle the resource selection problem. There are three types of supervised methods: query classification, resource classification and learning to rank methods. Dai et al. \cite{r:19} proposed a learning to rank method where they trained an SVMrank model based on features such as query-independent information, term-based statistics and CSI-based features. Another method is LTRRS \cite{r:20}, which uses term matching, CSI based, and topical relevance features. These features are combined to build a LambdaMART model that directly optimizes resources ranking list based on the nDCG metric to rank resources. The main drawbacks of these group of methods are that they require extensive feature engineering and heavily rely on domain knowledge, which makes it difficult to capture diverse feature aspects of queries and resources.

Combination-based methods combine some of the approaches mentioned above. The method ECOMSVZ \cite{r:21} combines a number of strategies to rank resources according to a given query, namely i) topical closeness between a query and resources, ii) estimation of a resource popularity from online sources, and iii) query expansion. The authors of \cite{r:22} present a method that aggregates all three strategies proposed in \cite{r:21} and ranks resources by giving more weights to small (specialized) resources.

\begin{figure*}
  \includegraphics[width=\textwidth]{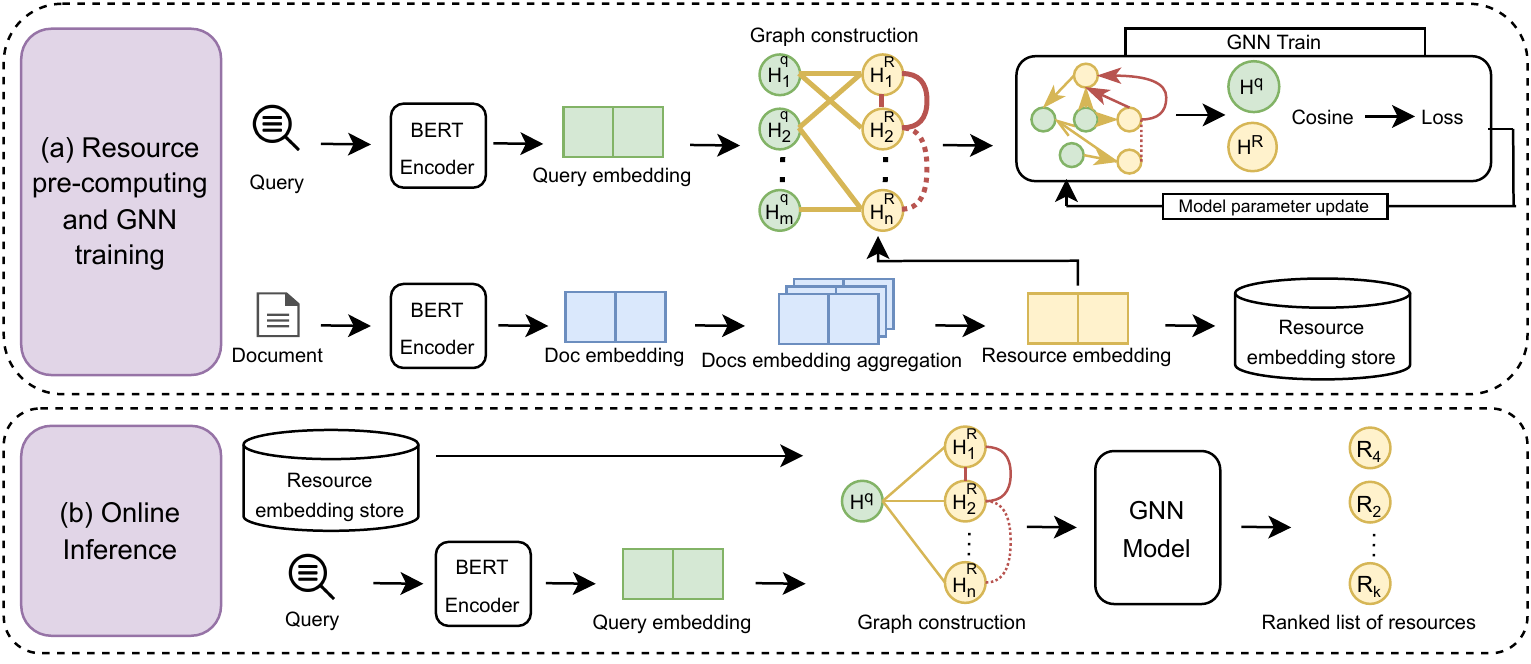}
  \caption{The overall architecture of the proposed method. Documents and queries are fed into a BERT encoder one at a time for semantic feature extraction. After obtaining resource embeddings from the aggregated document embeddings, resource embeddings are stored for the usage in the inference phase. A heterogeneous graph is constructed 
  using query-document pairs where the edges in yellow and red denote query-resource
and resource-resource edges, respectively. GNN is trained based on the constructed graph. In the inference phase, after obtaining an embedding from a BERT encoder for a user query, a graph is built based on the stored resource embeddings. Finally, a trained GNN model outputs a ranked listed of resources.}
  \label{fig:architecture}
\end{figure*}

The work most related to ours is that of Hong et al \cite{r:28}. It proposes a joint probabilistic classification model for resource selection. The model utilizes logistic regression to estimate the probability of relevance of information sources in a joint manner by considering both the evidence of individual sources and their relationships. However, logistic regression has a number of assumptions and limitations for modeling relations between variables \cite{r:32}. Some limitations include considering dependent variable structure, observation independence, absence of multicollinearity, linearity of independent variables and log odds, and large sample size \cite{r:33}. In practice, if the data used do not satisfy the assumed conditions, the use of logistic regression model may 
incur significant errors \cite{r:34}.

We propose a new learning to rank approach that utilizes a PTLM and GNN to address the limitations listed above. Our method takes into account both the semantics of query and resource as well as the  
query-resource and resource-resource relationships. 
First, we obtain embeddings from a pre-trained language model for queries and resources. Then, we construct a heterogeneous graph based on query-resource relevance judgement scores. To our knowledge, our approach is the first to employ a pre-trained language model BERT and GNN for the resource selection problem.
This gives us a richer resource representation, which in turn allows us to identify the most relevant resources for a given user query.

\section{Our Method}
\label{sec:method}
In this section, we describe FedGNN. We start with the task formulation.
The objective of the resource selection task is to output a ranked list of resources given a query representing an information need. Our approach falls into the learning to rank class of methods. 

Let $R=\{R_1,R_2,…,R_N\}$ be the resources to be ranked for a given user query $q$. Suppose the relevance scores of the resources are given as multi-level ratings $Z = \{l(R_1,q),l(R_2,q),…,l(R_N,q)\}$, where $l(R_i,q)\in\{r_1,…,r_N\}$ denotes the relevance score of $R_i$ for a user query $q$. Without loss of generality, we assume $l(R_i,q)\in\{0,1,…,N-1\}$ and name the corresponding relevance score as N-level ratings. If $l(R_i,q)>l(R_j,q)$, then $R_i$ should be ranked before $R_j$. Let $\mathcal{F}$ be the function class and $g \in \mathcal{F}$ be a ranking function. The goal of learning to rank is to learn the optimal ranking function from the training data by minimizing a certain loss function defined on the resources and their labels.

Since we do not know the optimal $g$, we use $\theta$ to denote the parameters that we need to learn toward finding (an approximation of) $g$. We propose to learn the optimal parameters $\theta$ of a function $g(\cdot; \theta)$ from the training data by minimizing a pointwise loss function defined as
\begin{equation}
    \mathcal{L}(R,q;\theta) = \sum^{n}_{i=1} \left(Cosine(g(R_i,q;\theta)) - l(R_i,q) \right)^2
\end{equation}
where $g$ represents the GNN model.

There is a counter-intuitive connection between ranking measures and loss functions in learning to rank methods. On the one hand, we learn a ranking function by minimizing a pointwise loss function. On the other hand, it is the ranking measures, like nDCG or nP@k,  that we use to evaluate the performance of the learned ranking function. Thus, one may wonder if the minimization of the loss functions can lead to the optimization of the ranking measures. A number of works looked into this question and proved that the minimization of these loss functions will lead to the maximization of the ranking measures \cite{r:52, r:53, r:54}. Relevant to our work, it has been proved that the regression and classification based losses used in the pointwise approach are upper bounds of (1 - nDCG) \cite{r:52, r:53}.


\subsection{Architecture Overview}
Figure \ref{fig:architecture} depicts the architecture of our proposed method. The method consists of the following five modules: (i) Dataset preprocessing, (ii) BERT encoder based feature extraction, (iii) Resource embedding generation from the aggregated documents embeddings, (iv) Graph construction, and (v) GNN based feature extraction with similarity score prediction.

Given a collection of search results (query-document pairs) with human labelled relevance judgement, we select top N documents per resource with the highest labelled judgements. Then, we input documents and queries one at a time into a BERT encoder for semantic feature extraction. After obtaining resource embeddings from the aggregated document embeddings, we store resource embeddings for the usage in the inference phase, and also we construct a heterogeneous graph according to the provided query-document pairs. In the training phase, the constructed heterogeneous graph is fed into GNN to extract the structural features for query and resource nodes. Then, we minimize Mean Squared Error (MSE) based on relevance judgement and computed Cosine similarity score between query and resource. In the inference phase, we obtain an embedding from a BERT encoder for a user query and build a graph based on the stored resource embeddings so that a query node connects to each resource node. Finally, a trained GNN model outputs a similarity score for each resource.

\subsection{Data Preprocessing}
The Data Preprocessing module is mainly responsible for the selection of sampled documents to represent each resource. Instead of considering all the sampled documents for every training query, the module takes only 
a subset of top-ranked documents over all queries. Specifically, we sort the documents based on their similarities with training queries, and select the N unique documents with the overall highest similarities, per each resource. Then, we use a BERT encoder to obtain semantic representations for queries and top N documents for each resource.

\subsection{Query and Resource Representations} \label{bert_representation}
To represent a document, we use its title concatenated with its body text, while to represent a query, we use query text. We adopt a feature-based approach to obtain query and resource representations using a BERT encoder. There are major computational benefits to pre-compute an expensive representation of the training data once and then run many experiments with cheaper task-specific learning to rank models on top of this representation \cite{r:37}. Hence, we first construct a heterogeneous graph from the representations and store resource embeddings for the further usage in the inference phase, then we train a model with a 
GNN algorithm.

BERT encoder takes a sequence of tokens as input and outputs a sequence of their corresponding final embeddings, i.e., one embedding for each token. Different from other feature-based approaches \cite{r:39}, we do not add special token [CLS] at the beginning of the token sequence. Following the original BERT \cite{r:37,r:40}, each query and document is converted to a token sequence using the BERT tokenizer, as in Equation \ref{eq:1}.
\begin{equation}
    \label{eq:1}
    \{e_1,...,e_i\}=BertTokenizer(\{t_1,...,t_i\})
\end{equation}
where $\{e_1,...,e_i\}$ and $\{t_1,...,t_i\}$ represent a token sequence and the words in a text (i.e., query or document), respectively. Then, we feed the token sequence into a BERT encoder to obtain a token representation
\begin{equation}
    \{E_1,...,E_i\}=BERT(\{e_1,...,e_i\})
\end{equation}
where $\{E_1,...,E_i\}$ is the last hidden state of BERT output. The last hidden state contains one embedding for each token (in other words, for each word in a text) where the size of a single embedding is 768. To obtain a final representation for an input (query or document), we calculate the mean of token embeddings.
\begin{equation}
    H = \frac{1}{K} \sum^{K}_{i=1} E_i
\end{equation}
where $K$ is the number of tokens in a sequence. Formally, we denote feature representations for queries and documents as $H_i^q \in R^{d}$ and $H_i^D \in R^{d}$, where $d$ is the embedding dimensionality, while $q$ and $D$ denote query and document embeddings, respectively. 
Once the feature embeddings for the selected top N documents are obtained, feature embedding for the $j$-th resource is calculated by taking the arithmetic mean of the top N documents per each resource:
\begin{equation}
    H_j^R = \frac{1}{N} \sum^{N}_{i=1} H_i^D
\end{equation}

\subsection{Graph Construction}

We build a heterogeneous graph, denoted as $\mathcal{G}=(\mathcal{V},\mathcal{E},\mathcal{R})$, where $\mathcal{V}$ and $\mathcal{E}$ are a set of nodes and edges, respectively. $\mathcal{R}$ represents a set of edge relations. For example, a tuple $\langle q_i,query-resource,R_j \rangle \in \mathcal{E}$ denotes that a query node $q_i$ and a resource node $R_j$ are connected by the relation $query-resource \in \mathcal{R}$. There are two types of edge relations $\mathcal{R}=\{query$-$resource$, $resource$-$resource\}$. For convenience, we denote $query$-$resource$ and $resource$-$resource$ edge relations as $qr$ and $rr$, respectively. Query and resource node embeddings are initialized with the output from the BERT encoder. 

From the given $query$-$document$ relevance judgement scores, we calculate the $query$-$resource$ relevance score by summing the $query$-$document$ relevance scores (grouped by query and resource) multiplied by a normalization coefficient $\alpha$:
\begin{equation}
    \hat{z}_{ij} = \alpha * \sum^{n}_{k=1} r_{k}
\end{equation}
where $\hat{z}_{ij}$ is the ground truth relevance score between the $i$-th query and the $j$-th resource, $r_{k}$ is the ground truth relevance judgement score between the query and the $k$-th document, and $n$ denotes the number of data points. We also utilize ground truth relevance scores as the weights on edges of the $query$-$resource$ relation type. For convenience, we use $E^{qr}=\hat{z}$ to denote weights on edges of the $query$-$resource$ relation type.

Weights on edges of the $resource$-$resource$ relation type are derived from the Cosine similarity function between two resource embeddings: 
\begin{equation}
    E^{rr}_{ij} = Cosine\left(H_i^R,H_j^R\right)
\end{equation}
where $H_i^R$ and $H_j^R$ are embeddings of the $i$-th and the $j$-th resources, respectively.
Figure \ref{fig:architecture} illustrates a constructed heterogeneous graph, where yellow lines represent $query$-$resource$ edges and red lines are of the $resource$-$resource$ edge type.

\subsection{The GNN Module}

In this paper, we adopt Relational Graph Convolutional Network (R-GCN) \cite{r:24} as a representative from the GNN family because it is designed to cope with heterogeneous type of graphs. We use R-GCN to capture the structural nature of the resource selection problem. Next, we will briefly introduce R-GCN.

\subsubsection{Relational Graph Convolutional Network}
R-GCN \cite{r:24} is an extension of the Graph Convolutional Network (GCN), which performs the information exchange in a multi-relational graph. As discussed in Section \ref{bert_representation}, we leverage BERT encoder to initialize $H^0$ for query and resource nodes. The representation of a node for the $l$-th layer is defined by:
\begin{equation}
    \bar{H}_i^{(l+1)} = \sigma \left( \sum_{r \in \mathcal{R}} \sum_{j \in \mathcal{N}_i^r} E^r_{ij} W_r^{l} H_j^{l} + W_0^{l} H_i^{l} \right)
\end{equation}
where $\mathcal{R}=\{qr,rr\}$ is the set of edge relation types and $N_i^r$ denotes the set of neighbors of node $v_i$ based on edges of relation type $r \in \mathcal{R}$. $W_r^l$ and $W_0^l$ are learnable parameter matrices, and $E^r_{ij}$ is a scalar weight on the edge of type $r$ between the $i$-th and the $j$-th nodes. To simplify, we can write such a message passing process as:
\begin{equation}
    \bar{H}^{(l+1)} = GNN(H^{l};\theta)
\end{equation}
where $\theta$ is the learning parameter and $\bar{H}$ is the final representations for query and resource nodes.  

\subsection{Final Prediction and Loss Function}

To predict a relevance score between a query and a resource, we employ the Cosine similarity:
\begin{equation}
    z_{ij} = Cosine \left(\bar{H}_i^q, \bar{H}_j^R \right)
\end{equation}
where $\bar{H}_i^q$ and $\bar{H}_j^R$ are the final embedding vectors for the $i$-th query $q_i$ and the $j$-th resource $R_j$ obtained from GNN, respectively.

Finally, we adopt the pointwise ranking objective as a loss function. The idea of pointwise ranking function is to measure the distance between the ground truth relevance score and the predicted relevance score. For this, we utilize Mean Square Error (MSE) loss:
\begin{equation}
    \mathcal{L} = \sum^{n}_{i=1} \left( \hat{z}_{ij} - z_{ij} \right)^2
\end{equation}
where $z_{ij}$ and $\hat{z}_{ij}$ are predicted and ground truth relevance scores, respectively, and $n$ denotes the number of data points.  

\section{Experiments}
\label{sec:experiments}
To evaluate the effectiveness of our proposed method in resource selection, in this section, we validate this method and compare it against a number of baseline models. First, we describe baseline models and real-world public datasets used in our comprehensive experimental study and provide the experimental details. Finally, we present empirical results with discussions and analysis.

\subsection{Baselines}
To examine the performance of the proposed model, we take all four categories of resource selection approaches as baselines, including lexicon-based (Taily \cite{r:9}), sample-based (Rank-S \cite{r:27}, ReDDE \cite{r:11}, CRCS \cite{r:15} and KBCS \cite{r:10}), supervised (L2R \cite{r:19} and Jnt \cite{r:28}) and combination-based (SSLTS \cite{r:22} and ECOMSVZ \cite{r:21}). These baseline methods are briefly described below:

\noindent\textbf{KBCS} \cite{r:10} is one of the recently proposed methods in the sample-based category that has the best performance among methods of this category.\\
\noindent\textbf{L2R} \cite{r:19} is a learning to rank method that trains SVMrank based on features such as query-independent information, term-based statistics and sample-document features.\\
\noindent\textbf{ECOMSVZ} \cite{r:21} is combination-based method that achieves one of the best results for selecting relevant resources 
in TREC 2014.
It combines a number of strategies to rank resources for a given query, namely i) topical closeness between a query and resources, ii) estimation of a resource popularity from online sources, and iii) query expansion.\\
\noindent\textbf{SSLTS} \cite{r:22} is also a combination-based method and consists of all three strategies proposed in \cite{r:21} and ranks resources by giving more weights to small (specialized) resources.\\
\noindent\textbf{Taily} \cite{r:9} is a method in the lexicon-based category. It models a query’s score distribution in each resource as a Gamma distribution and selects resources with highly scored documents in the tail of the distribution.\\
\noindent\textbf{Rank-S} \cite{r:27} is a sample-based algorithm from the SHiRE family of methods, where the scores of retrieved documents from the CSI are decayed exponentially and then treated as votes for the resources the documents were sampled from.\\
\noindent\textbf{ReDDE} \cite{r:11} is another effective and commonly used 
sample-based resource selection method.\\
\noindent\textbf{CRCS} \cite{r:15} is a sample-based method and, like ReDDE, it runs the query on a CSI and, in contrast to ReDDE, ranks resources according to the positions of retrieved documents in the CSI ranking.\\
\noindent\textbf{Jnt} \cite{r:28} is a learning to rank model. It employs a joint probabilistic classification model that estimates the probabilities of relevance in a joint manner by considering relationships among resources.

\subsection{Comparison Methodology}
To compare with the baselines fairly, we compare the experimental results of our method with those that were reported in the literature directly based on the same benchmark datasets and the same evaluation metrics. More specifically, if the reported experimental result for a baseline B is based on dataset D and evaluation metric M, we compare with B using the same D and the same M. An advantage of this comparison approach is that we do not need to implement the baselines ourselves. The down side of this approach is that since different datasets and different evaluation metrics were used in reported literature, the comparison is less than uniform, i.e., we are not able to compare with all baselines using the same datasets and the same evaluation metrics uniformly.

\subsection{Evaluation Metrics}
As mentioned earlier, to facilitate fair comparison, we adopt the same evaluation metrics used in the baseline methods. 
The following evaluation metrics will be used in our experimental study:
\begin{itemize}[leftmargin=*]
    \item Precision at $k$ or P@$k$. Precision is the number of relevant retrieved items with respect to the total number of retrieved items. P@10 is used to compare with baselines L2R \cite{r:19}, KBCS \cite{r:10}, ReDDE \cite{r:11}, CRCS \cite{r:15}, Rank-S \cite{r:27}, Taily \cite{r:9} and Jnt \cite{r:28}.
    \item Normalized precision at $k$ or nP@$k$. This metric measures the relevance scores of the top ranked k resources, normalized by the relevance scores of the best possible k resources for the given query. nP@\{1,5\} is used to compare with baselines SSLTS \cite{r:22} and ECOMSVZ \cite{r:21}.
    \item Normalized Discounted Cumulated Gain at $k$ or nDCG@$k$. This metric allows different degrees of relevance of resources to be considered. nDCG also discounts the value of a relevant resource when a resource is ranked lower. nDCG@\{10,20\} is used to compare with SSLTS \cite{r:22} and ECOMSVZ \cite{r:21}.
\end{itemize}

P@$k$ and nDCG@$k$ are widely used evaluation metrics in information retrieval and resource selection. 
The precision-oriented metrics at lower ranks such 10 or 30 (e.g., P@10 and nDCG@30) are expected to perform 
better than recall-based metrics at deeper ranks (e.g., P@100 and nDCG@100) in the resource selection problem because resource selection returns a small portion of resources \cite{r:23}. Moreover, since users are usually interested in the first 10 results, shallow ranked relevant results 
are expected to directly 
meet users' satisfaction.

\subsection{Datasets}
We run our experiments on the following two widely used benchmark datasets. The baselines were evaluated on those datasets, too.

\noindent\textbf{ClueWeb09-B}\footnote{http://www.lemurproject.org/clueweb09}  is a subset of the full web collection ClueWeb09. The dataset was created to support research on information retrieval and related human language technologies, and it contains around 50 million English webpages. This collection is divided into different resources to evaluate resource selection methods (see below for more details). \\
\noindent\textbf{FedWeb Greatest Hits} \cite{r:26} is a large test collection designed to support research in web information retrieval including resource selection in federated search. The collection contains two datasets FedWeb13 and FedWeb14, which consist of 157 and 149 resources, respectively. 
We use the FedWeb14 dataset because it has about 3 million sampled documents. The dataset also has 275 information needs (queries), and out of those only 50 queries are judged. 

According to L2R \cite{r:19}, experiments were conducted with the ClueWeb09-B dataset which was divided into 123 resources. The relevance scores were obtained from TREC Web Track 2009-2012\footnote{https://trec.nist.gov/data/webmain.html}. Another baseline that used the same dataset in experiments is KBCS \cite{r:10}. They divided it into 100 resources, and evaluation queries were acquired from TREC Web Track 2009. Other two baseline methods SSLTS \cite{r:22} and ECOMSVZ \cite{r:21} used the FedWeb14 dataset, and for evaluation, judged queries were utilized.

\begin{table}
    \setlength\tabcolsep{2.5pt}
    \centering
    \caption{Dataset statistics}
    \label{tab:data-stat}
    \scalebox{0.8}{
        \begin{tabular}{ccccccc}
            \toprule
                \multirow{2}{*}{\makecell{\thead{Dataset}}} & \multirow{2}{*}{\makecell{\thead{\# of \\ resources}}} & \multirow{2}{*}[2mm]{\makecell{\thead{Total \\ \# of \\ docs (K)}}} & \multicolumn{3}{c}{\thead{Resource (K)}} & \multirow{2}{*}[2mm]{\makecell{\thead{Compared \\ with \\ baselines}}} \\ \cline{4-6} 
                    &   &   & \makecell{\thead{Min}} & \makecell{\thead{Max}} & \makecell{\thead{Avg}} &   \\
            \midrule
                ClueWeb09-B & 100 & 50,220 & 63 & 1,691 & 502 & \makecell{KBCS, ReDDE, \\ CRCS, Rank-S}\\
            \hline
                ClueWeb09-B & 123 & 50,220 & 32 & 734 & 408 & \makecell{L2R, ReDDE, \\ Rank-S, Taily, Jnt}\\
            \hline
                FedWeb14 & 149 & 187.7 & .022 & 2.7 & 1.2 & SSLTS, ECOMSVZ\\
            \bottomrule
        \end{tabular}
    }
\end{table}

Since our proposed learning to rank algorithm requires ground truth relevance scores to train a model, we leverage only judged portion of the queries from the FedWeb14 dataset for both training and evaluation. For the ClueWeb09-B dataset with 123 resources setup, we use TREC Web Track 2009-2012 for training and evaluation. TREC Web Track 2009 is used for the ClueWeb09-B dataset with 100 resources setup to train and evaluate a model.

\begin{table}
    \setlength\tabcolsep{3pt}
    \centering
    \caption{Query set statistics}
    \label{tab:query-stat}
    \scalebox{0.8}{
        \begin{tabular}{cccc}
            \toprule
                \makecell{\thead{Query set}} & \makecell{\thead{\# of \\ queries}} & \makecell{\thead{Folds}} & \makecell{\thead{Compared with \\ baselines}}\\
            \midrule
                \makecell{Web Track \\ 2009} & 50 & 5 & \makecell{KBCS, ReDDE, \\ CRCS, Rank-S}\\
            \hline
                \makecell{Web Track \\ 2009-2012} & 200 & 10 & \makecell{L2R, ReDDE, \\ Rank-S, Taily, Jnt}\\
            \hline
                FedWeb14 & 50 & 5 & \makecell{SSLTS, \\ ECOMSVZ}\\
            \bottomrule
        \end{tabular}
    }
\end{table}

Table \ref{tab:data-stat} summaries the statistics of the datasets. For the FedWeb14 dataset, we consider only judged documents. In addition, the last column “Comparison with baselines” lists the baseline methods that are used to compare with our method using a given dataset.

\begin{table}
    \setlength\tabcolsep{3pt}
    \centering
    \caption{Graphs statistics for three datasets}
    \label{tab:graph-stat}
    \scalebox{0.8}{
        \begin{tabular}{ccccc}
            \toprule
             \makecell{\thead{Datasets}} & \makecell{\thead{\# of \\ resource \\ nodes}} & \makecell{\thead{\# of \\ query \\ nodes}} & \makecell{\thead{\# of \\ query-resource \\ edges}}  & \makecell{\thead{\# of \\ resource-resource \\ edges}}\\
            \midrule
                \makecell{ClueWeb09-B} & 123 & 200 & 4046 & 7503 \\
            \hline
                \makecell{ClueWeb09-B} & 100 & 50 & 1618 & 4950 \\
            \hline
                FedWeb14 & 149 & 50 & 4286 & 11026 \\
            \bottomrule
        \end{tabular}
    }
\end{table}

\begin{table}
    \setlength\tabcolsep{3pt}
    \parbox{.45\linewidth}{
        \centering
        \caption{Search accuracy comparison for ClueWeb09-B with 123 resources}
        \label{tab:exp-clueweb123}
        \scalebox{0.8}{
            \begin{tabular}{ccc}
                \toprule
                    \multirow{2}{*}{\makecell{Methods}} & \multicolumn{2}{c}{P@10} \\ \cline{2-3}
                                                        & \makecell{T=4} & \makecell{T=8}    \\
                \midrule
                    ReDDE & 0.355 & 0.363 \\
                    Rank-S & 0.350 & 0.259 \\
                    Taily & 0.346 & 0.260 \\
                    Jnt & 0.370 & 0.367 \\
                    L2R & 0.374 & 0.377 \\
                    FedGNN & \textbf{0.398} & \textbf{0.407} \\
                \bottomrule
            \end{tabular}
        }
        
    }
    \hfill
    \parbox{.45\linewidth}{
        \setlength\tabcolsep{3pt}
        \centering
        \caption{Search accuracy comparison for ClueWeb09-B with 100 resources}
        \label{tab:exp-clueweb100}
        \scalebox{0.8}{
            \begin{tabular}{ccc}
                \toprule
                    \multirow{2}{*}{\makecell{Methods}} & \multicolumn{2}{c}{P@10} \\ \cline{2-3}
                                                        & \makecell{T=4} & \makecell{T=6}    \\
                \midrule
                    ReDDE & 0.244 & 0.252 \\
                    CRCS & 0.236 & 0.252 \\
                    Rank-S & 0.316 & 0.365 \\
                    KBCS & 0.370 & 0.400 \\
                    FedGNN & \textbf{0.418} & \textbf{0.429} \\
                \bottomrule
            \end{tabular}
        }
    }
\end{table}

Table \ref{tab:query-stat} gives query sets that are used to train and evaluate the proposed model. It also shows cross-validation settings we followed for each query set. In cross-validation, we partition queries. For example, FedWeb14 has 50 queries. For five-fold cross-validation, the 50 queries are partitioned into 5 subsets; in each run, four subsets are used for training and one subset is used for testing. The results are then averaged over the five runs. Graph statistics for three datasets are presented in Table \ref{tab:graph-stat}

\subsection{Implementation Details}
To obtain the initial embeddings for document titles and queries, we utilize $BERT_{base}$ architecture that encodes input text to a 768-dimensional dense vector space. After aggregating the top 10 documents’ embeddings to represent resource nodes, we leverage the DGL\footnote{https://www.dgl.ai/} python package to build a graph and train a GNN model. We implemented the proposed method in PyTorch. We trained FedGNN model with Adam optimizer \cite{r:29} and followed cross-validation settings shown in Table \ref{tab:query-stat}. The results reported in this paper are based on the following parameter settings: RelGraphConv aggregator: pool, HeteroGraphConv aggregator: mean, hidden dropout rate: 0.3 (FedWeb14), input feature size: 768, hidden size: 768, output size: 768. All experiments were conducted on a Debian server equipped with CPU: Intel(R) Core(TM) i9-10940X, GPU: NVIDIA Quadro RTX 6000 24GB, RAM: 64GB. To simulate a distributed information retrieval environment, we followed same search engine setup parameters given in the baseline papers. When resource selection decides which resource to retrieve, then for each query, top 1000 documents are returned from each resource and merged directly.

\subsection{Experimental Results}
There are three sets of experiments to verify the effectiveness of our method. One set compares ours with L2R \cite{r:19} and its baselines Taily \cite{r:9}, Rank-S \cite{r:27}, ReDDE \cite{r:11}, and Jnt \cite{r:28} on the document level as shown in Table \ref{tab:exp-clueweb123}. Table \ref{tab:exp-clueweb100} reports the results of the second set of experiments, showing a comparison with KBCS \cite{r:10} as well as its baselines ReDDE \cite{r:11}, CRCS \cite{r:15}, and Rank-S \cite{r:27} on the document level. The results of the final set of experiments are reported in Table \ref{tab:exp-fedweb14}, comparing the effectiveness of our method with those of SSLTS \cite{r:22} and ECOMSVZ \cite{r:21}.

\begin{table}
    \setlength\tabcolsep{3pt}
    \centering
    \caption{Search accuracy comparison for FedWeb14}
    \label{tab:exp-fedweb14}
    \scalebox{0.8}{
        \begin{tabular}{ccccc}
            \toprule
                Methods & nDCG@10 & nDCG@20 & nP@1 & nP@5\\
            \midrule
                SSLTS (Market share)&0.55&0.66&0.38&0.48\\
                ECOMSVZ&0.712&0.624&0.535&0.604\\
                FedGNN&\textbf{0.794}&\textbf{0.821}&\textbf{0.76}&\textbf{0.788}\\
            \bottomrule
        \end{tabular}
    }
\end{table}

The empirical results show that FedGNN outperforms the baselines significantly in all metrics suggesting that the combination of a BERT and GNN based method can improve the effectiveness of resource selection. Specifically, according to the results given in Table \ref{tab:exp-clueweb123}, our model outperforms the best baseline in its group, L2R, by $6.4\%$ and $8\%$ on P@10 for top 4 and 8 resources, respectively. On dataset ClueWeb09-B with 100 resources (reported in Table \ref{tab:exp-clueweb100}), FedGNN outperforms the best baseline in its group, KBCS, by $13\%$ and $7.3\%$ on nP@10 for top 4 and 6 resources, respectively. The final experiment with the dataset FedWeb14, the results provided in Table \ref{tab:exp-fedweb14} show that FedGNN outperforms the best baseline in its group, ECOMSVZ, by $11.5\%$ on nDCG@10, by $42\%$ on nP@1, and by $30.4\%$ on nP@5. On nDCG@20, the best baseline is SSLTS, and FedGNN outperforms this baseline by $24.4\%$. 

\begin{figure}
    \centering
    \includegraphics[scale=0.5]{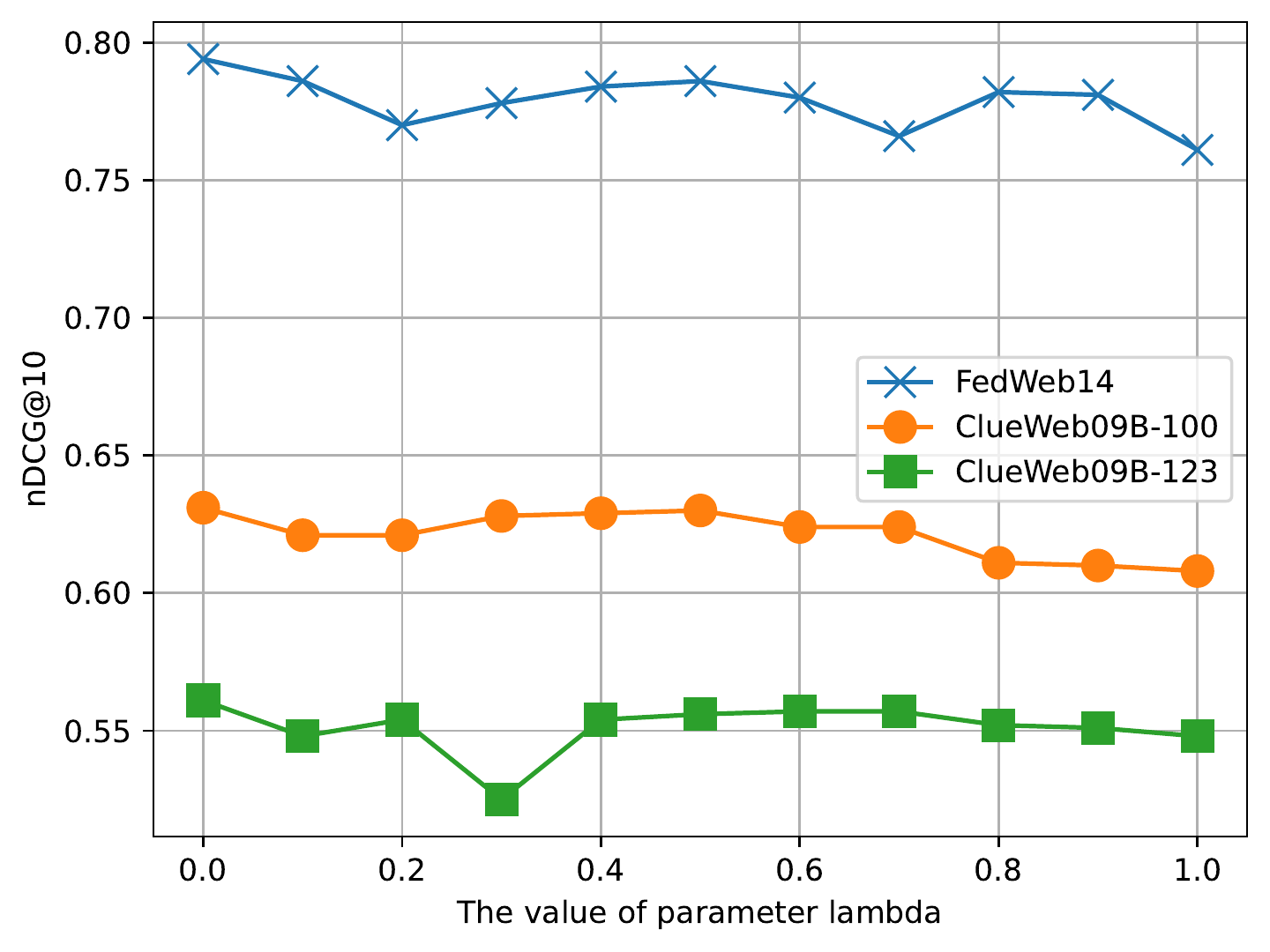}
     \captionof{figure}{The nDCG@10 measure of FedGNN for different $\lambda$}
    \label{fig:iden-lambda}
    \vspace{-4mm}
\end{figure}

One of the reasons that may have contributed to the above significant improvements is that our proposed method does not suffer from the "bigger gets bigger vote" problem. Specifically, Shokouhi et al \cite{r:15} reports that the bigger the collection size, the higher the chances are that it contains more relevant documents. Since sample-based methods are based on a centralized sample index (CSI), most of those methods are biased towards bigger resources by giving larger resources higher scores. On the other hand, our proposed method does not require CSI which is expensive to obtain. Our method needs a small number of documents from each resource, and it aggregates embeddings of documents to obtain a resource embedding. Thus, we leverage a single source of truth to represent a resource. In this way, resource discrimination by its size is avoided which may have contributed to the better effectiveness of our proposed method over the baselines. 

Another reason for the superiority of FedGNN is the interactive aggregation of features from BERT and GNN modules. It leads to the combination of the advantages of these two features to the greatest extent which guarantees a better effectiveness over the existing methods. More specifically, the baselines L2R, KBCS, SSLTS and ECOMSVZ utilize many term-based statistical features which do not capture semantic information. It is well-known that a user query may have many meanings in different contexts, and hence, capturing semantic information is crucial in federated search. Different from the baselines, our model can not only consider semantic information of a user query, but also integrates the structural information between query and resource nodes. Query and resource nodes in a constructed graph get useful messages from their two-hop neighbors, and it leads to better ranking results \cite{r:55, r:56, r:57}.

\subsection{Ablation Study}
To analyze the usefulness of each component of FedGNN, we perform the following experiments. 

\subsubsection{Effectiveness of the GNN}
Our model 
leverage the aggregated features from BERT and GNN. Compared with the original BERT, FedGNN integrates structural features in addition to semantic features. To validate the effectiveness of GNN, we design an experiment to compare FedGNN with a model without GNN. Essentially, the model employs Cosine similarity function to get a relevance score between query and resource embeddings obtained from BERT encoder. We name the model as FedBERT. Moreover, to validate the effectiveness of Relational Graph Convolutional Network (R-GCN) \cite{r:24} used in our model, we substitute that with the Heterogeneous Graph Transformer (HGT) \cite{r:25} algorithm; we name it FedHGT. Table \ref{tab:effect-gnn} illustrates the experimental results. Compared with using BERT alone, FedGNN achieves significantly better results on the three datasets. This confirms the necessity of adding GNN to capture structural features. The empirical results also elucidate that R-GCN performs better than HGT, which implies R-GCN generalizes better on unseen queries 
in the resource selection problem. This is the chief contribution of our paper.

\begin{table}
\renewcommand{\arraystretch}{1.2}
\setlength\tabcolsep{4pt}
\centering
\caption{Ablation study of the effectiveness of the GNN}
\label{tab:effect-gnn}
\scalebox{0.8}{
    \begin{tabular}{ccccccc}
    \toprule
    \multirow{2}{*}{Methods} & \multicolumn{2}{c}{FedWeb14} & \multicolumn{2}{c}{ClueWeb09B-100} & \multicolumn{2}{c}{ClueWeb09B-123} \\ \cline{2-7} 
                             & nDCG@10 & nP@10 & nDCG@10 & nP@10 & nDCG@10 & nP@10 \\ \midrule
    FedBERT & 0.088 & 0.107 & 0.261 & 0.405 & 0.343 & 0.575 \\ 
    FedHGT & 0.71 & 0.668 & 0.592 & 0.812 & 0.38 & 0.637 \\ 
    FedGNN & \textbf{0.794} & \textbf{0.808} & \textbf{0.631} & \textbf{0.856} & \textbf{0.561} & \textbf{0.788} \\ \bottomrule
    \end{tabular}
}
\vspace{-4mm}
\end{table}

\subsubsection{The impact of resource-resource edges}
We investigate the effect of resource-resource edges on the ranking. To obtain resource-resource edges, we calculate Cosine similarity between resource nodes, and conduct an experiment to determine the optimal value of parameter $\lambda$. $\lambda$ is a threshold that controls the number of resource-resource edges based on their similarity scores. For example, $\lambda=0.5$ means we keep the edges with similarity scores at least $0.5$.

\begin{table*}
\renewcommand{\arraystretch}{1.2}
\setlength\tabcolsep{4pt}
\centering
\caption{Ablation study of the effect of BERT encoder}
\label{tab:effect-bert}
\scalebox{0.8}{
    \begin{tabular}{ccccccccc}
    \toprule
    \multirow{2}{*}{Methods} & \multicolumn{2}{c}{FedWeb14} & \multicolumn{2}{c}{ClueWeb09B-100} & \multicolumn{2}{c}{ClueWeb09B-123} & \multicolumn{2}{c}{Average} \\ \cline{2-9} 
                             & nDCG@10 & nP@10 & nDCG@10 & nP@10 & nDCG@10 & nP@10 & nDCG@10 & nP@10 \\ \midrule
    FedGNN-WV & 0.789 & \textbf{0.812} & 0.57 & 0.721 & 0.551 & 0.782 & 0.637 & 0.772 \\ 
    FedGNN-DV & 0.738 & 0.781 & 0.593 & 0.807 & 0.542 & 0.773 & 0.624 & 0.787 \\ 
    FedGNN-FT & \textbf{0.806} & \textbf{0.812} & 0.577 & 0.75 & 0.547 & 0.751 & 0.643 & 0.771 \\ 
    FedGNN & 0.794 & 0.808 & \textbf{0.631} & \textbf{0.856} & \textbf{0.561} & \textbf{0.788} & \textbf{0.662} & \textbf{0.817} \\ \bottomrule
    \end{tabular}
}
\end{table*}

In the experiment, we increase the value of $\lambda$ from 0 to 1 with a fixed step of $0.1$. Figure \ref{fig:iden-lambda} illustrates the empirical results from the experiment conducted on the FedGNN model with the FedWeb14, ClueWeb09B-100 and ClueWeb09B-123 datasets. Note that when $\lambda=1$, it will essentially exclude all resource-resource edges, leaving the graph consisting of only query-resource type of edges.

The experimental results indicate that the presence of resource-resource type of edges in a graph increases the overall ranking performance, which proves that structure-based semantic similarities are important in the resource selection problem. On FedWeb14 dataset, nDCG@10 is increased by 4\% when $\lambda=0.0$ is compared to $\lambda=1.0$. While on datasets ClueWeb09B-100 and ClueWeb09B-123, nDCG@10 is increased by 3.8\% and 2.4\%, respectively. Moreover, the evaluation metric nDCG is maximized at the same value of parameter $\lambda$ for all three datasets: $\lambda=0.0$. Resource-resource relationships, at such a value of $\lambda$ when the evaluation metric is maximized, are important and may carry crucial insights into an overall structure of a graph. Hence, $\lambda$ is set to the above identified value (i.e., 0.0) in our all analysis with the baselines.

\subsubsection{Effect of BERT encoder}
In this section, we explore the effect of BERT encoder to the overall ranking. To validate, we substitute BERT encoder with some of the most popular language models like Word2Vec \cite{r:41}, Doc2Vec \cite{r:42}, FastText \cite{r:43}, and compare them with our model. Specifically, we pre-train Word2Vec, Doc2Vec and FastText language models for each dataset (e.g. FedWeb14, ClueWeb09B-100, ClueWeb09B-123). Then, we develop three models namely FedGNN-WV, FedGNN-DV, FedGNN-FT that utilize Word2Vec, Doc2Vec and FastText, respectively. Since Word2Vec and FastText generate an embedding on a word level, we average word embeddings to obtain a representation on a document level.

Table \ref{tab:effect-bert} shows that Word2Vec and FastText perform slightly better than BERT on nP@10 metric over FedWeb14 dataset. FastText also outperforms BERT based model only by $1.5\%$ on nDCG@10. However, BERT outperforms the best baselines on average over three datasets by $3\%$ and $4\%$ on nDCG@10 and nP@10, respectively. The reasons for the superiority of BERT based model is its ability in representing the contextualized embedding. Different from other language models, BERT considers the order of words in a sentence which leads to syntactic and semantic understanding of a sentence \cite{r:37}. This validates the importance of BERT encoder in our model. 

\subsubsection{Effect of different hidden layers of BERT encoder}
$BERT_{BASE}$ architecture has 12 transformer blocks (a.k.a hidden layers or layers). Authors of BERT \cite{r:37} consider some methods that generate embeddings from different hidden layers. In this experiment, we aim to determine a method that increases an overall ranking of resources by generating a better contextualized representation for queries and resources. The following methods are considered in the experiment: [CLS] token of the last hidden layer (CLS), concatenation of the last 4 hidden layers (C4LH), sum of the last 4 hidden layers (S4LH), last hidden layer (LHCLS), last hidden layer without special tokens like [CLS] (LH-CLS).

\begin{table}
    \renewcommand{\arraystretch}{1.2}
    \setlength\tabcolsep{3pt}
    \centering
    \caption{Ablation study of the effect of different hidden layers of BERT encoder on nDCG@10 metric}
    \label{tab:effect-lh-bert}
    \scalebox{0.8}{
        \begin{tabular}{cccccc}
            \toprule
                Datasets & CLS & C4LH & S4LH & LHCLS & LH-CLS \\
            \midrule
                FedWeb14 & 0.773 & 0.742 & 0.775 & 0.778 & \textbf{0.794} \\
                ClueWeb09B-100 & 0.62 & 0.617 & 0.616 & 0.614 & \textbf{0.631} \\
                ClueWeb09B-123 & 0.547 & 0.559 & 0.537 & 0.56 & \textbf{0.561} \\
            \bottomrule
        \end{tabular}
    }
\end{table}

\begin{table}
\renewcommand{\arraystretch}{1.2}
\setlength\tabcolsep{2.5pt}
\centering
\caption{Ablation study on document title and body text}
\label{tab:effect-title-body}
\scalebox{0.8}{
    \begin{tabular}{ccccccc}
    \toprule
    \multirow{2}{*}{Datasets} & \multicolumn{2}{c}{Title} & \multicolumn{2}{c}{Body} & \multicolumn{2}{c}{Title+Body} \\ \cline{2-7} 
                             & nDCG@10 & nP@10 & nDCG@10 & nP@10 & nDCG@10 & nP@10 \\ \midrule
    FedWeb14 & 0.779 & 0.796 & 0.775 & 0.783 & \textbf{0.794} & \textbf{0.808} \\ 
    ClueWeb09B-100 & 0.617 & 0.833 & 0.62 & 0.82 & \textbf{0.631} & \textbf{0.856} \\ 
    ClueWeb09B-123 & 0.558 & 0.776 & 0.548 & 0.774 & \textbf{0.561} & \textbf{0.788} \\ \bottomrule
    \end{tabular}
}
\vspace{-4mm}
\end{table}

Table \ref{tab:effect-lh-bert} shows the FedGNN results trained on five embedding generation methods. The results are reported on nDCG@10 metric on three datasets. The results indicate that the LH-CLS method outperforms the best methods by $1.29\%$ on average. Hence, we utilize an embedding obtained from the last hidden layer of BERT encoder without special tokens. 

\subsubsection{Effect of document title and body text}
We designed an experiment to validate that the concatenation of title and body text gives a better representation of a document that leads to a higher overall ranking of resources. 

Table \ref{tab:effect-title-body} shows empirical results of three FedGNN models trained on three different document attributes: Title, Body and Title+Body. Input texts larger than 512 tokens are truncated. The experimental results show that a model trained on Title+Body attribute outperforms the best results trained on either Title or Body on average by $1.4\%$ and $1.94\%$ on nDCG@10 and nP@10, respectively.

\section{Conclusion}
\label{sec:conclusions}

In this paper, we tackled the resource selection problem in federated search. We proposed a learning to rank with GNN approach that addresses resource representation and resource selection problems at the same time. The proposed approach takes advantage of both a pre-trained language model and a heterogeneous graph for learning and ranking resources based on the given query. Our approach consists of five modules: data processing module, a pre-trained language module for extracting semantic information from queries and documents, a module for obtaining a resource representation, a heterogeneous graph construction module and a GNN module for capturing structural dependencies between queries and resources as well as among resources. Comprehensive experiments on two benchmark datasets demonstrate the superiority of the proposed method compared with the state-of-the-art approaches. Specifically, our method outperforms the state-of-the-art by 6.4\% to 42\% on various performance metrics. Moreover, the ablation studies prove the effectiveness of each module of the FedGNN model.

\begin{acks} 
This work was supported in part by the U.S. NSF 1838145 and 2107213 grants,  Temple University office of the Vice President for Research 2022 Catalytic Collaborative Research Initiative Program and a gift from NVIDIA Corporation.
\end{acks}

\bibliographystyle{ACM-Reference-Format}
\bibliography{references}

\appendix

\end{document}